\documentclass[prX,showpacs,twocolumn]{revtex4}

\usepackage{graphicx}

\usepackage{amsmath,srcltx}
\usepackage{latexsym}
\usepackage{amssymb}    
\usepackage{amsfonts}

\newcommand{\be}{\begin{equation}}
\newcommand{\ee}{\end{equation}}
\newcommand{\ba}{\begin{eqnarray}}
\newcommand{\ea}{\end{eqnarray}}

\def\a{\alpha}
\def\b{\beta}
\def\d{\delta}
\def\e{\epsilon}

\def\vf{\varphi}
\def\g{\gamma}
\def\h{\eta}
\def\j{\psi}

\def\m{\mu}
\def\n{\nu}

\def\p{\pi}

\def\r{\rho}
\def\s{\sigma}

\def\D{\Delta}
\def\F{\Phi}

\def\ca{{\cal A}}

\def\cj{{\cal J}}

\newcommand{\ov}{\overline}

\newcommand{\aand}{\;\;\;\mbox{and}\;\;\;}
\newcommand{\pa}{\partial}

\newcommand{\pari}{\stackrel{{P}}\longrightarrow}

\newcommand{\rae}{\rangle}
\newcommand{\lae}{\langle}
\def\sl#1{\rlap{\hbox{$\mskip 1 mu /$}}#1}
\def\Sl#1{\rlap{\hbox{$\mskip 3 mu /$}}#1}
\def\SL#1{\rlap{\hbox{$\mskip 4 mu /$}}#1}
\def\I{\leavevmode\hbox{\small1\kern-3.8pt\norfeynmpmalsize1}}

\begin{document}
\title{Electron-polaron--electron-polaron bound states in mass-gap graphene-like planar quantum electrodynamics: $s$-wave bipolarons}
\author{O.M. Del Cima}
\email{oswaldo.delcima@ufv.br}
\affiliation{Universidade Federal de Vi\c cosa (UFV),\\
Departamento de F\'\i sica - Campus Universit\'ario,\\
Avenida Peter Henry Rolfs s/n - 36570-900 -
Vi\c cosa - MG - Brazil.}

\author{E.S. Miranda}
\email{emerson.s.miranda@ufv.br}
\affiliation{Universidade Federal de Vi\c cosa (UFV),\\
Departamento de F\'\i sica - Campus Universit\'ario,\\
Avenida Peter Henry Rolfs s/n - 36570-900 -
Vi\c cosa - MG - Brazil.}

\begin{abstract}

A Lorentz invariant version of a mass-gap graphene-like planar quantum electrodynamics, the parity-preserving $U(1)\times U(1)$ massive QED$_3$, exhibits attractive interaction in low-energy electron-polaron--electron-polaron $s$-wave scattering, favoring quasiparticles bound states, the $s$-wave bipolarons.

\end{abstract}
\maketitle

\section{Introduction}
The seminal works by Deser, Jackiw, Templeton and Schonfeld \cite{deser-jackiw-templeton-schonfeld} have attracted attention to the quantum electrodynamics in three space-time dimensions (QED$_3$) in view of its potentiality as theoretical foundation for quasi-planar condensed matter phenomena, such as high-$T_{\rm c}$ superconductors \cite{high-Tc}, quantum Hall effect \cite{quantum-hall-effect}, topological insulators \cite{topological-insulators}, topological superconductors \cite{topological-superconductors} and graphene \cite{graphene}. Since then, the planar quantum electrodynamics has been widely studied in many physical configurations, namely, small (perturbative) and large (non perturbative) gauge transformations, Abelian and non-Abelian gauge groups, fermions families, odd and even under parity, compact space-times, space-times with boundaries, curved space-times, discrete (lattice) space-times, external fields and finite temperatures. 

The pure graphene \cite{graphene} monolayer is a gapless (massless gap graphene) bidimensional system which behaves like a half-filling semimetal with its charge carriers (quasiparticles) being described by massless charged Dirac fermions. However, for practical applications like transistors a gap (mass-gap) graphene \cite{mass-gap-graphene} is more appropriate, and such a mass-gap effect is observed in pure monolayer graphene on substrates \cite{hagues}. Electron-electron interactions (electron pairing) in graphene \cite{electron-pairing} include electron polarons (electron-phonon) \cite{electron-phonon} scattering processes \cite{polarons}, where this quasiparticle, the polaron, which is formed by a bound state of electron (or hole) and phonon, was first introduced by Landau \cite{landau}.

The proposed issue in this work about the possibility of $s$-wave bipolarons emerging from the parity-preserving $U(1)\times U(1)$ massive QED$_3$ -- a mass-gap graphene-like \cite{mass-gap-graphene} planar quantum electrodynamics -- is presented as follows. Initially, the model defined by its discrete and continuous symmetries is introduced and, since the interactions are nonconfining -- the vector mesons, the photon and the N\'eel quasiparticle, are massive -- the asymptotic states for the fermions (electron polarons) are established. Hereafter, in the low-energy limit, the $s$- and $p$-wave M{\o}ller ($e^-$-polaron--$e^-$-polaron) scattering amplitudes are computed and their respective interaction potentials obtained and analysed. However, from this analysis, it was found conditions on the parameters which, in spite of the $p$-wave scattering potential still remains repulsive, the $s$-wave interaction potential becomes attractive. The latter shall favour $e^-$-polaron--$e^-$-polaron bound states -- provided the attractive $s$-wave scattering potential satisfies necessary conditions \cite{chadan-khuri-martin-wu,kato,newton-seto,bargmann} -- giving rise to the $s$-wave bipolarons condensates \cite{polarons}.

\section{The model}
The Lorentz invariant version of mass-gap graphene-like planar quantum electrodynamics, the parity-even $U(1)_A\times U(1)_a$ massive QED$_3$, is defined by the action:
\ba
&& S=\int{d^3 x} \biggl\{-{1\over4}F^{\m\n}F_{\m\n} - {1\over4}f^{\m\n}f_{\m\n} + 
\m\e^{\m\r\n}A_\m\pa_\r a_\n + \nonumber\\
&&+~i {\ov\j_+} {\Sl D} \j_+ + i {\ov\j_-} {\Sl D} \j_- - m({\ov\j_+}\j_+ - {\ov\j_-}\j_-) + \nonumber\\
&&-~\frac{1}{2\a}(\pa^\m A_\m)^2 - \frac{1}{2\b}(\pa^\m a_\m)^2 \biggr\}~,\label{action}
\ea
where ${\SL D}\j_\pm\equiv (\sl\pa + ie\Sl{A} \pm ig\sl{a})\j_\pm$, and any object 
${\SL X}\equiv X^\m\g_\m$. The coupling constants $e$ and $g$ are dimensionful, with mass dimension $\frac12$, and, $m$ and $\m$ are mass parameters with mass 
dimension $1$. Also, $F_{\m\n}=\pa_\mu A_\nu - \pa_\n A_\m$ and 
$f_{\m\n}=\pa_\mu a_\nu - \pa_\n a_\m$, are the field strengths associated to the electromagnetic field ($A_\m$) and the N\'eel gauge field ($a_\m$), respectively, $\j_+$ and $\j_-$ are 
two kinds of fermions -- each of them describing electron polarons (electron-phonon) and hole polarons (hole-phonon) quasiparticles -- where the $\pm$ subscripts are related to their spin sign \cite{binegar}, and 
the gamma matrices are $\g^\m=(\s_z,-i\s_x,i\s_y)$. 

\subsection{The symmetries: parity and $U(1) \times U(1)$}
The CPT-even action (\ref{action}) is invariant under: 
\begin{enumerate}
\item parity symmetry ($P$):        
\ba
x_\m \!\! & \pari &\!\! x_\m^P=(x_0,-x_1,x_2)~,\nonumber\\
\j_\pm \!\! & \pari &\!\! \j_\pm^P=-i\g^1\j_\mp~,~\ov\j_\pm \pari \ov\j_\pm^P=i\ov\j_\mp\g^1~,\nonumber \\
A_\mu \!\! & \pari &\!\! A_\mu^P=(A_0,-A_1,A_2)~,\nonumber\\
a_\mu \!\! & \pari &\!\! a_\mu^P=(-a_0,a_1,-a_2)~.
\label{xp}
\ea
\item gauge $U(1)_A \times U(1)_a$ symmetry ($\delta_{\rm g}$):
\ba 
&&\delta_{\rm g} \psi_\pm(x)=i[\theta(x)\pm\omega(x)]\psi_\pm(x)~,~~\nonumber \\
&&\delta_{\rm g} \ov{\psi}_\pm(x)=-i[\theta(x) \pm \omega(x)]\ov{\psi}_\pm(x)~,\nonumber \\
&&\delta_{\rm g} A_{\mu}(x)=- \frac{1}{e}\,\partial_{\mu}\theta(x)~,\nonumber \\ 
&&\delta_{\rm g} a_{\mu}(x)=- \frac{1}{g}\,\partial_{\mu}\omega(x)~.\label{gaugetransf}
\ea
\end{enumerate}

\subsection{The spectrum: degrees of freedom, spin, masses and charges}
The free Dirac equations associated to $\j_+$ and $\j_-$, which stem from the action (\ref{action}), read: 
\be
(i{\sl\pa} - m)\j_+=0 \aand (i{\sl\pa} + m)\j_-=0~,\label{dirac}
\ee
So, by expanding the operators $\j_+$ and $\j_-$ in terms of the $c$-number plane wave solutions of the Dirac equations, with operator-valued amplitudes, $a_+$, $b_+$, $a_-$ and $b_-$ (annihilation operators), and $a_+^\dag$, $b_+^\dag$, $a_-^\dag$ and $b_-^\dag$ (creation operators):
\ba
 \j_+(x)=\int\frac{d^2\vec k}{(2\p)^2}\frac{m}{k^0}&&\{a_+(k) u_+(k)e^{-ikx} + \nonumber\\ 
 &&+~ b_+^\dag(k) v_+(k)e^{ikx}\}~, \label{psi+}\\
 \j_-(x)=\int\frac{d^2\vec k}{(2\p)^2}\frac{m}{k^0}&&\{a_-(k) u_-(k)e^{-ikx} + \nonumber\\
 &&+~ b_-^\dag(k) v_-(k)e^{ikx}\}~, \label{psi-}
\ea
where $\ov\j_\pm=\j_\pm^\dag \g^0$. Consequently, from (\ref{dirac}) and 
(\ref{psi+})-(\ref{psi-}), by assuming $p^\m=(E,p_x,p_y)$, the wave functions, $u_+$, $v_+$, $u_-$ and $v_-$, are given by:
\ba
u_+(p)=\frac{({\sl p}+m)}{\sqrt{2m(E+m)}}u_+(m,\vec 0)~, \nonumber \\ 
v_+(p)=\frac{(-{\sl p}+m)}{\sqrt{2m(E+m)}}v_+(m,\vec 0)~, \\
u_-(p)=\frac{(-{\sl p}+m)}{\sqrt{2m(E+m)}}u_-(m,\vec 0)~, \nonumber \\ 
v_-(p)=\frac{({\sl p}+m)}{\sqrt{2m(E+m)}}v_-(m,\vec 0)~, 
\ea 
satisfying the following conditions:
\ba
&&\ov{u}_+(p)u_+(p)=1 \aand \ov{v}_+(p)v_+(p)=-1~, \label{norma+} \\ 
&&\ov{u}_-(p)u_-(p)=-1 \aand \ov{v}_-(p)v_-(p)=1~,   \label{norma-}
\ea
where 
\ba 
&&u_+(m,\vec 0)=\left(\begin{array}{c}
1  \\
0
\end{array}\right) 
\aand 
v_+(m,\vec 0)=\left(\begin{array}{c}
0  \\
1
\end{array}\right) , \\
&&u_-(m,\vec 0)=\left(\begin{array}{c}
0  \\
1
\end{array}\right)
\aand 
v_-(m,\vec 0)=\left(\begin{array}{c}
1 \\
0
\end{array}\right) ,
\ea  
are the momenta space solutions of the Dirac equations at the particle rest-frame, 
$p^\m=(m,\vec{0})$. 
The microcausality conditions for $\j_+$ and $\j_-$: 
\be
\big\{\j_\pm(x),\j_\pm^\dag(y)\big\}_{x^0=y^0}=\d^2(\vec{x}-\vec{y})~,
\ee  
together with the Dirac equations (\ref{dirac}) and the normalization conditions 
(\ref{norma+})-(\ref{norma-}), implies that:
\ba
&&\big\{a_\pm(k),a_\pm^\dag(p)\big\}=(2\p)^2\frac{k^0}{m} \d^2(\vec{k}-\vec{p})~,\\
&&\big\{b_\pm(k),b_\pm^\dag(p)\big\}=(2\p)^2\frac{k^0}{m} \d^2(\vec{k}-\vec{p})~,
\ea
where all other anticommutators vanish and, for the vacuum state $|0\rangle$, 
$a_\pm(k)|0\rangle=b_\pm(k)|0\rangle=0$.

The quantum operators associated to space-time ($SO(1,2)$) symmetry and internal ($U(1)_A \times U(1)_a$) symmetry, spin ($S$), electric charge ($Q_\pm$) and N\'eel (chiral) charge ($q_\pm$), are 
\ba
S\!\!&=&\!\!\frac{1}{2}\s_z ~, \label{S}\\
Q_\pm \!\!&=&\!\! - e\int d^2\vec{x} :\j_\pm^{\dag}(x)\j_\pm(x): \label{Q+-}\\
      \!\!&=&\!\! - e\int\frac{d^2\vec k}{(2\p)^2}\frac{m}{k^0}\{a_\pm^\dag(k) a_\pm(k) - 
      b_\pm^\dag(k) b_\pm(k)\} ~, \nonumber\\
q_\pm \!\!&=&\!\! \mp g\int d^2\vec{x} :\j_\pm^{\dag}(x)\j_\pm(x): \label{q+-}\\
      \!\!&=&\!\! \mp g\int\frac{d^2\vec k}{(2\p)^2}\frac{m}{k^0}\{a_\pm^\dag(k) a_\pm(k) - 
      b_\pm^\dag(k) b_\pm(k)\} ~, \nonumber
\ea
respectively, with their action upon the asymptotic fermion (antifermion) states with spin up and spin down, 
$|f_{\uparrow}^-\rae$ ($|f_{\uparrow}^+\rae$) and $|f_{\downarrow}^-\rae$ ($|f_{\downarrow}^+\rae$): 
\ba
&S|f_{\uparrow}^-\rae = +\frac{1}{2}|f_{\uparrow}^-\rae ~,~~ 
S|f_{\downarrow}^+\rae = -\frac{1}{2}|f_{\downarrow}^+\rae ~, \nonumber \\ 
&S|f_{\downarrow}^-\rae = -\frac{1}{2}|f_{\downarrow}^-\rae ~,~~ 
S|f_{\uparrow}^+\rae = +\frac{1}{2}|f_{\uparrow}^+\rae ~; \\
&Q_+|f_{\uparrow}^-\rae = - e|f_{\uparrow}^-\rae ~,~~ 
Q_+|f_{\downarrow}^+\rae = + e|f_{\downarrow}^+\rae ~, \nonumber \\
&Q_-|f_{\downarrow}^-\rae = - e|f_{\downarrow}^-\rae ~,~~ 
Q_-|f_{\uparrow}^+\rae = + e|f_{\uparrow}^+\rae ~; \\
&q_+|f_{\uparrow}^-\rae = - g|f_{\uparrow}^-\rae ~,~~ 
q_+|f_{\downarrow}^+\rae = + g|f_{\downarrow}^+\rae ~, \nonumber \\
&q_-|f_{\downarrow}^-\rae = + g|f_{\downarrow}^-\rae ~,~~ 
q_-|f_{\uparrow}^+\rae = - g|f_{\uparrow}^+\rae ~;
\ea
where
\ba
|f_{\uparrow}^-\rae = a_+^\dag(k)|0\rae ~,&~~  |f_{\downarrow}^+\rae = b_+^\dag(k)|0\rae ~, \\ 
|f_{\downarrow}^-\rae = a_-^\dag(k)|0\rae ~,&~~ |f_{\uparrow}^+\rae = b_-^\dag(k)|0\rae ~, 
\ea
which means that, $a_+^\dag$ ($a_-^\dag$) creates a spin-up (spin-down) fermion (electron polaron) and  $b_+^\dag$ ($b_-^\dag$) creates a spin-down (spin-up) antifermion (hole polaron). Moreover, from the results above, for any fermion or antifermion (spin up or down) quantum state ${|\j\rae}$, it is verified that
\be
S {|\j\rae} = -\frac{1}{2g}~ q_{\pm} {|\j\rae}~, \label{spin-chiralcharge}
\ee
which proves the correlation among spin and chiral charge (see TABLE \ref{table1}).

\begin{table}
\begin{tabular}{|c|p{1,2cm}|p{1cm}|p{1cm}|c|c|} 
\hline
{state} & {$\;\,$ wave \newline function} & {electric charge} & {$\,$chiral charge} & {spin} & {quasiparticle}  \\
\hline\hline
{$|f_{\uparrow}^-\rae$} & ~~~~$u_+$  & ~~{$-e$}   & ~~{$-g$}  & $+\frac{1}{2}$ & {electron polaron} \\
\hline 
{$|f_{\downarrow}^-\rae$} & ~~~~$u_-$  & ~~{$-e$}   & ~~{$+g$}  & $-\frac{1}{2}$ & {electron polaron} \\
\hline
{$|f_{\downarrow}^+\rae$} & ~~~~$v_+$  & ~~{$+e$}   & ~~{$+g$}  & $-\frac{1}{2}$ & {hole polaron} \\
\hline
{$|f_{\uparrow}^+\rae$} & ~~~~$v_-$  & ~~{$+e$}   & ~~{$-g$}   & $+\frac{1}{2}$ & {hole polaron} \\
\hline
\end{tabular}
\caption[]{The quasiparticles electric charges, chiral charges and spin.}\label{table1}
\end{table}

In the low-energy limit (Born approximation), the two-particle scattering potential is given by the 
Fourier transform of the two-particle $t$-channel scattering amplitude (direct scattering) \cite{sakurai}. 
However, so as to compute the scattering amplitudes, use has been made of the propagators. Hence, switching off the coupling constants ($e$ and $g$), the tree-level propagators in momenta space, for all the fields, read: 
\ba
&&\D_{++}(k)=i\frac{\sl{k}-m}{k^2-m^2}~,~~\D_{--}(k)=i\frac{\sl{k}+m}{k^2-m^2}~;\label{propk++--}\\
&&\D^{\m\n}_{AA}(k)=
-i\biggl\{ \frac{1}{k^2-\m^2}\biggl(\h^{\m\n}-\frac{k^\m k^\n}{k^2}\biggr) + 
\frac{\a}{k^2}\frac{k^\m k^\n}{k^2} \biggr\}~, \label{propkAA}\nonumber\\
&&\D^{\m\n}_{aa}(k)=
-i\biggl\{ \frac{1}{k^2-\m^2}\biggl(\h^{\m\n}-\frac{k^\m k^\n}{k^2}\biggr) + 
\frac{\b}{k^2}\frac{k^\m k^\n}{k^2} \biggr\}~, \label{propkaa}\nonumber\\
&&\D^{\m\n}_{Aa}(k)=\D^{\m\n}_{aA}(k)=\frac{\m}{k^2(k^2-\m^2)}\e^{\m\r\n}k_\r~. \label{propkAa}
\ea
From the propagators above, $\D_{++}$, $\D_{--}$, $\D^{\m\n}_{AA}$, $\D^{\m\n}_{aa}$ and  $\D^{\m\n}_{Aa}$, the spectrum and the tree-level unitarity of the model can be be analyzed by coupling 
them to external currents, 
$\cj_{\F_i}=(\cj_+,\cj_-,\cj^{\m}_A,\cj^{\m}_a)$, compatible with the symmetries of the model, where the current-current transition amplitudes in momentum space are written as:
$\ca_{\F_i\F_j} = \cj_{\F_i}^*(k) \lae \F_i(k) \F_j(k) \rae \cj_{\F_j}(k)$.
Then, by taking the imaginary part of the residues of the current-current 
amplitudes, $\ca_{\F_i\F_j}$, at the poles, it can be probed the necessary conditions for unitarity -- positive imaginary part of the residues of the transition amplitudes, $\Im{\rm Res}~\ca_{\F_i\F_j}>0$, as a consequence of the $S$-matrix be unitary -- at the tree-level and the counting of the degrees of freedom described by the fields, $\F_i=(\j_+,\j_-,A_\m,a_\m)$. 
In summary, it has been concluded \cite{del_cima-miranda} that the two kind of fermions, $\j_+$ and 
$\j_-$, hold two massive degrees of freedom with mass $m$ -- the electron-polaron $|f_{\uparrow}^-\rae$ ($u_+$) and the hole-polaron $|f_{\downarrow}^+\rae$ ($v_+$) associated to the spinor $\j_+$, and the electron-polaron $|f_{\downarrow}^-\rae$ ($u_-$) and the hole-polaron $|f_{\uparrow}^+\rae$ ($v_-$) associated to the spinor $\j_-$. Also, the vector fields, the electromagnetic field ($A_\m$) and 
the N\'eel gauge field ($a_\m$), carry each one two massive degrees of freedom with mass $\m$, moreover,   
it shall be noticed that the single massless mode in model, displayed in $\D^{\m\n}_{Aa}$, does not propagate, it decouples. From the results presented above, it can be concluded that the the parity-preserving $U(1)\times U(1)$ massive QED$_3$ is free from tachyons and ghosts at the classical
level. Nevertheless, to have full control of the unitarity at tree-level, it is still necessary to study the behaviour of the scattering
cross sections in the limit of high center of mass energies, by analyzing the Froissart-Martin bound \cite{froissart-martin-bound}.

\section{The M{\o}ller scattering}
In order to calculate the scattering amplitudes, it remains the vertex Feynman rules associated to the interaction vertices $-e{\ov\j_\pm} \Sl{A} \j_\pm$ and 
$\mp g{\ov\j_\pm} \sl{a} \j_\pm$: $\Upsilon_{\pm \pm}^\m\!\!=\!\!ie\g^\m$ and 
$\upsilon_{\pm \pm}^\m\!\!=\!\!\pm ig\g^\m$, respectively. 

\begin{figure}[h]
\centering
\setlength{\unitlength}{1,0mm}
\includegraphics[width=5.5cm,height=2cm]{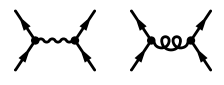}
\caption[]{$e^-$-polaron--$e^-$-polaron (M{\o}ller) $t$-channel scattering mediated  
by electromagnetic ($A_\mu$) and N\'eel ($a_\mu$) quantum fields.}
\label{moller}
\end{figure}

The $t$-channel $e^-$-polaron--$e^-$-polaron M{\o}ller scattering amplitudes mediated 
by the electromagnetic and the N\'eel quanta (see FIG. \ref{moller}) are given by:
\ba
&&\!\!\!\!\!\!\!\!-i\mathcal{M}_{\pm A \mp} = \nonumber\\
&&\!\!\!\!\!\!\!\!\ov{u}_\pm(p_1')[\Upsilon_{\pm\pm}^\m]u_\pm(p_1)\D_{\m\n}^{AA}(k)\ov{u}_\mp(p_2')[\Upsilon_{\mp\mp}^\n]u_\mp(p_2)~,\label{amplicovariante1}\\ 
&&\!\!\!\!\!\!\!\!-i\mathcal{M}_{\pm a \mp} = \nonumber\\
&&\!\!\!\!\!\!\!\!\ov{u}_\pm(p_1')[\upsilon_{\pm\pm}^\m]u_\pm(p_1)\D_{\m\n}^{aa}(k)\ov{u}_\mp(p_2')[\upsilon_{\mp\mp}^\n]u_\mp(p_2)~,\label{amplicovariante2}\\
&&\!\!\!\!\!\!\!\!-i\mathcal{M}_{\pm A \pm} = \nonumber\\
&&\!\!\!\!\!\!\!\!\ov{u}_\pm(p_1')[\Upsilon_{\pm\pm}^\m]u_\pm(p_1)\D_{\m\n}^{AA}(k)\ov{u}_\pm(p_2')[\Upsilon_{\pm\pm}^\n]u_\pm(p_2)~,\label{amplicovariante3}\\ 
&&\!\!\!\!\!\!\!\!-i\mathcal{M}_{\pm a \pm} = \nonumber\\
&&\!\!\!\!\!\!\!\!\ov{u}_\pm(p_1')[\upsilon_{\pm\pm}^\m]u_\pm(p_1)\D_{\m\n}^{aa}(k)\ov{u}_\pm(p_2')[\upsilon_{\pm\pm}^\n]u_\pm(p_2)~.\label{amplicovariante4}
\ea
Furthermore, in the center of mass (CM) reference frame, the three-momenta configuration of the two scattered fermions, $p_1$, $p_2$, $p_1'$ and $p_2'$, so as the momentum transfer, $k$, are fixed as
\ba
&&\!\!\!\!\!\!\!\!\!\!\!\!p_1 = (E,p,0) ~,~~ p_1' = (E,p\cos\vf,p\sin\vf)~;\label{momenta1}\\
&&\!\!\!\!\!\!\!\!\!\!\!\!p_2 = (E,-p,0) ~,~~ p_2' = (E,-p\cos\vf,-p\sin\vf)~;\label{momenta2}\\
&&\!\!\!\!\!\!\!\!\!\!\!\!k = p_1 - p_1' = (0,p(1 - \cos\vf),-p \sin\vf) = (0, {\mathbf{k}})~, \label{momentak}
\ea 
where $\vf$ is the CM scattering angle, defined as the angle among the
directions in the CM frame of the two incoming (initial state) and
outgoing (final state) fermions.

The total $s$- and $p$-wave M{\o}ller scattering amplitudes can now be derived from the partial ones (\ref{amplicovariante1})-(\ref{amplicovariante4}) in the low-energy approximation, $\mathcal{M}_{s}$ 
($\mid\uparrow\rae +\! \mid\downarrow\rae \rightarrow\;  \mid\uparrow\rae +\! \mid\downarrow\rae$) and 
$\mathcal{M}_{p}$ 
($\mid\uparrow\rae +\! \mid\uparrow\rae \rightarrow\; \mid\uparrow\rae +\! \mid\uparrow\rae$ or 
$\mid\downarrow\rae +\! \mid\downarrow\rae \rightarrow\; \mid\downarrow\rae +\! \mid\downarrow\rae$), where, by assuming the momenta configuration above (\ref{momenta1})-(\ref{momentak}), it follows that:
\ba
&&\mathcal{M}_{s}=\frac{1}{{\mathbf{k}}^2 + \m^2}\left(e^2-g^2\right)~, \label{Ms}\\
&&\mathcal{M}_{p}=\frac{1}{{\mathbf{k}}^2 + \m^2}\left(e^2+g^2\right)~. \label{Mp}
\ea

\subsection{Scattering potentials}
In the low-energy (nonrelativistic) limit, the two-particle interaction potential, in the Born approximation,  is nothing but the two-dimensional Fourier transform of the lowest-order two-particle 
$\mathcal{M}$ scattering amplitude:
\be
\mathcal{V}(r)=
\int\frac{d^2{\mathbf{k}}}{(2\p)^2}~\mathcal{M}~ e^{i{\mathbf{k}}\cdot{\mathbf{r}}}~. \label{V}
\ee

Accordingly to the Born approximation (\ref{V}), the electron-polaron--electron-polaron $s$- and $p$-wave scattering potentials, mediated by the photon and the N\'eel quasiparticle, read: 
\ba
&&\mathcal{V}_{s}(r)=\frac{1}{2\p}\left(e^2-g^2\right)K_0(\m r)~, \label{Vs}\\
&&\mathcal{V}_{p}(r)=\frac{1}{2\p}\left(e^2+g^2\right)K_0(\m r)~. \label{Vp}
\ea
Thereafter, it can be concluded from (\ref{Vp}) that, regardless the values of the electromagnetic and the chiral coupling constants -- $e$ and $g$, respectively -- the $e^-$-polaron--$e^-$-polaron interaction in 
$p$-wave state ($\mid\uparrow\rae +\!\! \mid\uparrow\rae$ or $\mid\downarrow\rae +\!\! \mid\downarrow\rae$) is always repulsive. Nevertheless, from (\ref{Vs}), it shall be stressed about the possibility of attractive $e^-$-polaron--$e^-$-polaron interaction in $s$-wave state ($\mid\uparrow\rae +\! \mid\downarrow\rae$) provided $g^2>e^2$. In this case, where $g^2>e^2$, the $s$-wave interaction potential $\mathcal{V}_{s}(r)$ is attractive, 
\be
\mathcal{V}_{s}(r)=-\frac{1}{2\p}\left(g^2-e^2\right)K_0(\m r) ~, \label{Vs<0}
\ee
however, this is not a sufficient condition for the existence of bound states.

\subsection{Bound states}
Beyond the attractive nature, provided that $g^2>e^2$, of $s$-wave interaction potential (\ref{Vs<0}) it has to be weak in the sense of Kato \cite{kato},
\be
\int_0^{\infty}d\r~ \r[1 + |{\rm ln}(\r)|]|\mathcal{V}(\r)| < \infty ~,~~ \r=\m r~, \label{kato}
\ee
so as to satisfy the Newton-Set\^o and the Bargmann bounds \cite{newton-seto,bargmann}, which guarantee bound states and establish an upper bound for their number ($N^0$) for vanishing angular momentum ($l=0$):
\ba
&&\!\!\!\!\!\!\!\!\!\!\!\!\!\!\!\!N^0 < 1 + \nonumber\\
&&\!\!\!\!\!\!\!\!\!\!\!\!\!\!\!\!+~\frac{1}{2}\frac{\displaystyle \left(\frac{2\m_{\rm r}}{\hbar^2}\right)^2 \int_0^{\infty}\int_0^{\infty} \r\r'
\bigg\vert{\rm ln}(\frac{\r}{\r'})\bigg\vert|\mathcal{V}(\r)||\mathcal{V}(\r')|d\r d\r'}
{\displaystyle \m^2\left(\frac{2\m_{\rm r}}{\hbar^2}\right)\int_0^{\infty} \r|\mathcal{V}(\r)|d\r} ~; 
\label{newton-seto}
\ea
and upper bound for their number ($N^l$) for nonvanishing angular momentum ($l\neq 0$):
\be
N^l < \frac{1}{2l}\left(\frac{2\m_{\rm r}}{\hbar^2}\right)\frac{1}{\m^2} \int_0^{\infty} \r|\mathcal{V}(\r)|d\r ~, 
\label{bargmann}
\ee
respectively, where $\hbar=1$ and $\m_{\rm r}=\frac{m}{2}$ is the $e^-$-polaron--$e^-$-polaron reduced mass.

\begin{figure}[h]
\centering
\setlength{\unitlength}{1,0mm}
\includegraphics[width=6.5cm,height=4cm]{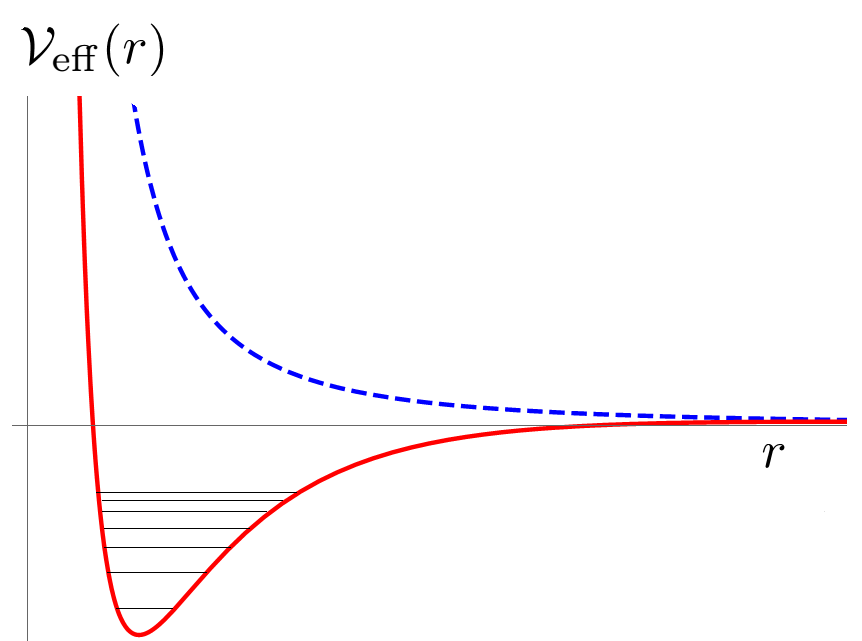}
\caption[]{The effective $e^-$-polaron--$e^-$-polaron interaction potential, 
which for the $s$-wave case, $C<0$, $\mathcal{V}_{\rm eff}(r)$ is attrative, if $0 \leq l \leq l_{\rm m}$ (solid line), and repulsive, if $l > l_{\rm m}$ (dashed line).}
\label{Veff}
\end{figure}

It has been proved elsewhere \cite{del_cima-franco-lima} that, whenever an interaction potential of the 
type $\mathcal{V}(r) = C K_0(\mu r)$ is attractive ($C<0$), it satisfies the following criteria: the weakness in the sense of Kato (\ref{kato}); 
the Newton-Set\^o bound (\ref{newton-seto}) for $l=0$; and the Bargmann bound (\ref{bargmann}) for all $l$ such that 
$l \leq l_{\rm m}=\lfloor \frac{\m_{\rm r}|C|}{\hbar^2\mu^2} \rfloor$ (where $\lfloor x \rfloor$ is the floor function of 
$x$). In the same manner, by means of the effective potential  
$\mathcal{V}_{\rm eff}(r)=\frac{\hbar^2(l^2-\frac{1}{4})}{2\mu_{\rm r}r^2} + \mathcal{V}(r)$ with $0 \leq l \leq l_{\rm m}$, 
it can be figured out that bound states arise (see FIG. \ref{Veff}). In addition to, it shall be stressed that these fulfilled conditions, (\ref{kato}), (\ref{newton-seto}) and (\ref{bargmann}), guarantee the existence of bound states for any kind of three-dimensional space-time model which exhibits scattering potential of the type 
$\mathcal{V}(r) = C K_0(\mu r)$ ($C<0$).

\section{Conclusions}
The Lorentz invariant parity-preserving $U(1)\times U(1)$ massive QED$_3$, a mass-gap graphene-like planar quantum electrodynamics model, at low-energy limit exhibits electron-polaron--electron-polaron scattering short range non confining potentials, similarly it can be concluded that the same behaviour takes place for hole-polaron--hole-polaron scatterings. The interactions among electron-polarons and hole-polarons are mediated by two massive vector mesons, the photon (electric charge source) and the N\'eel quasiparticle (chiral charge source), both stemming from the $U(1)_A\times U(1)_a$ gauge symmetry. It should be noticed that it was disclosed the correlation among the electron-polaron (hole-polaron) spin polarization and correspondent  chiral charge. At the tree-level, the absence of tachyons ($k^2<0$) and ghosts ($\lae\j|\j\rae<0$) in the model spectrum guarantees causality and unitarity, respectively, at this level. Notwithstanding, in order to complete the tree-level unitarity analysis, it remains to finish the  proof that the scattering cross sections in the limit of high center of mass energies respect the Froissart-Martin bound \cite{froissart-martin-bound}, but since ultraviolet problems are less critical in lower dimensional quantum field models, together with the fact that the four space-time dimensional QED (QED$_4$) \cite{sakurai} satisfies the Froissart-Martin bound, consequently this fulfilment shall be foreseen for parity-even $U(1)\times U(1)$ massive QED$_3$. Also, it shall be pointed out that for condensed matter systems like graphene, the quasiparticles (electron-polaron and hole-polaron) dynamics is in non relativistic regime, so ultraviolet unitarity upper bound violations should not be expected.

Bearing in mind hypothetical applications of the model presented here to graphene, or any other two dimensional system, the orders of magnitude of some theoretical parameters need to be established firstly, namely, a typical mass-gap in graphene is around meV \cite{mass-gap-graphene} whereas the low-energy limit for a condensed matter system is of eV order. In addition to that, the characteristic range of the two interactions, mediated by the both massive photon and the N\'eel quantum, shall be associated to the pair-coherence length measured in graphene, orders of magnitude in nm \cite{pair-coherence-length}. The mass-gap in graphene \cite{mass-gap-graphene}, besides of being more realistic, can be either achieved when pure graphene monolayer is settled on substrates \cite{hagues}, increasing its application range and improving device developments.

At the low-energy limit, the non relativistic electron-polaron--electron-polaron (or hole-polaron--hole-polaron) scattering potential, owing to photon and N\'eel quasiparticle short range exchanges, shows to be always repulsive (\ref{Vp}) for parallel ($p$-wave) electron-polaron (hole-polaron) spin polarizations ($\mid\uparrow\rae$+$\mid\uparrow\rae$ or $\mid\downarrow\rae$+$\mid\downarrow\rae$). Nevertheless, for electron-polaron--electron-polaron (or hole-polaron--hole-polaron) scatterings with antiparallel ($s$-wave) spin polarizations ($\mid\uparrow\rae$+$\mid\downarrow\rae$), the $s$-wave interaction potential (\ref{Vs<0}) might be attractive provided $e^-$($e^+$)-polaron--N\'eel-quasiparticle coupling strength ($|g|$) be stronger than the strength of $e^-$($e^+$)-polaron--photon coupling ($|e|$), $g^2>e^2$. Moreover, the $s$-wave attractive scattering potential (\ref{Vs<0}) satisfies the Kato condition \cite{kato}, the Newton-Set\^o and the Bargmann upper bounds \cite{newton-seto,bargmann}, indicating that $s$-wave bipolarons \cite{polarons} might stem from these electron-polaron--electron-polaron quasiparticles bound states \cite{del_cima-franco-lima}. The possible emergence of such a Cooper-type $e^-$-polaron--$e^-$-polaron condensate (bipolaron) directly calls the issue of superconductivity in graphene \cite{graphene-superconductor}, thus a deep investigation on that deserves special attention.

\subsection*{Acknowledgments} 
The authors thank O. Piguet, J.A. Helay\"el-Neto, D.H.T. Franco and J.M. Fonseca, as well as to the anonymous referee for helpful comments and suggestions. Special thanks are due to L.G. Rizzi. O.M.D.C. dedicates this work to his father (Oswaldo Del Cima, {\it in memoriam}), mother (Victoria M. Del Cima, {\it in memoriam}), daughter (Vittoria) and son (Enzo). CAPES-Brazil is acknowledged for invaluable financial help. 

\subsection*{Author contribution statement} 
All authors have been contributed equally.


\begin{references}

\bibitem{deser-jackiw-templeton-schonfeld} J.F. Schonfeld, Nucl. Phys. B185 (1981) 157; R. Jackiw and S. Templeton, Phys. Rev. D23 (1981) 2291; S. Deser, R. Jackiw and S. Templeton, Ann. Phys. (NY) 140 (1982) 372, Phys. Rev. Lett. 48 (1982) 975 and Ann. Phys. (NY) 281 (2000) 409. 

\bibitem{high-Tc} M. Franz, Z. Te\v{s}anovi\'c and O. Vafek, Phys. Rev. B66 (2002) 054535; I.F. Herbut, Phys. Rev. B66 (2002) 094504; H.R. Christiansen, O.M. Del Cima, M.M. Ferreira Jr and J.A. Helay\"el-Neto, Int. J. Mod. Phys. A18 (2003) 725.

\bibitem{quantum-hall-effect} R.B. Laughlin, Phys. Rev. Lett. 50 (1983) 1395; A.M.J. Schakel, Phys. Rev. D43 (1991) 1428; A. Raya and E.D. Reyes, J. Phys. A: Math. Theor. 41 (2008) 355401.
 
\bibitem{topological-insulators} M.Z. Hasan and C.L. Kane, Rev. Mod. Phys. 82 (2010) 3045; 
C.L. Kane and J.E. Moore, Physics World 24 (2011) 32; 
J.M. Fonseca, W.A. Moura-Melo and A.R. Pereira, J. Appl. Phys. 111 (2012) 064913. 
 
\bibitem{topological-superconductors} M. Leijnse and K. Flensberg, Semicond. Sci. Technol. 27 (2012) 124003; S. Yonezawa, arXiv:1604.07930v3, AAPPS Bulletin 26 No.3 (2016) 3; M. Sato and Y. Ando, arXiv:1608.03395v2, Rep. Prog. Phys. (to appear). 

\bibitem{graphene} V.P. Gusynin, V.A. Miransky and I.A. Shovkovy, Phys. Rev. Lett. 73 (1994) 3499; K.S. Novoselov, A.K. Geim, S.V. Morozov, D. Jiang, M.I. Katsnelson, I.V. Grigorieva, S.V. Dubonos, A.A. Firsov,  Nature 438 (2005) 197; V.P. Gusynin, S.G. Sharapov and J.P. Carbotte, Int. J. Mod. Phys. B21 (2007) 4611; R. Jackiw and S.-Y. Pi, Phys. Rev. Lett. 98 (2007) 266402;  R. Jackiw and S.-Y. Pi, Phys. Rev. B78 (2008) 132104; M.I. Katsnelson and K.S. Novoselov, Solid State Comm. 143 (2007) 3; A.H. Castro Neto, F. Guinea, N.M.R. Peres, K.S. Novoselov and A.K. Geim, Rev. Mod. Phys. 81 (2009) 109; A. Shytov, M. Rudner, N. Gu, M.I. Katsnelson and L. Levitov, Solid State Commun. 149 (2009) 1087; V.P. Gusynin, ``Graphene and Quantum Electrodynamics'', talk given at the International Jubilee Seminar ``Current problems in Solid State Physics'', November (2011), Kharkov, Ukraine; E.M.C. Abreu, M.A. De Andrade, L.P.G. De Assis, J.A. Helay\"el-
Neto, A.L.M.A. Nogueira and R.C. Paschoal, J. High Energy Phys. 1105 (2011) 001.  

\bibitem{mass-gap-graphene} D.V. Khveshchenko, Phys. Rev. Lett. 87 (2001) 246802; D.V. Khveshchenko and H. Leal, Nucl. Phys. B687 (2004) 323; L. Covaci and M. Berciu, Phys. Rev. Lett. 100 (2008) 256405; D.A. Siegel, C. Hwang, A.V. Fedorov and A. Lanzara, New J. Phys. 14 (2012) 095006; M.S. Fuhrer, Science  340 (2013) 1413; P. Kumar, R. Skomski, P. Manchanda, A. Kashyap and P.A. Dowben, Curr. Appl. Phys. 14 (2014) S136; A. Sharma, V.N. Kotov and A.H. Castro Neto, arXiv:1702.03551v1;

\bibitem{hagues} J.P. Hagues, Phys. Rev. B 86 (2012) 064302.

\bibitem{electron-pairing} G.Q. Hai, L. C\^andido, B.G.A. Brito and F.M. Peeters, J. Phys. Commun. 2 (2018) 035017.

\bibitem{electron-phonon} C. Chen, J. Avila, S. Wang, Y. Wang, C. Shen, R. Yang, B. Nosarzewski, T.P. Devereaux, G. Zhang and M.C. Asensio, Nano Lett. 18 (2018) 1082.

\bibitem{polarons} J.T. Devreese, Z. Phys. B 104 (1997) 601; J.T. Devreese and A.S. Alexandrov, Rep. Prog. Phys. 72 (2009) 066501; D.A. Siegel, C. Hwang, A.V. Fedorov and A. Lanzara, New J. Phys. 14 (2012) 095006;
J.C. Johannsen, S. Ulstrup, M. Bianchi, R. Hatch, D. Guan, F. Mazzola, L. Hornek{\ae}r, F. Fromm, C. Raidel, T.  Seyller and P. Hofmann, J. Phys.: Condens. Matter 25 (2013) 094001; C. Chen, J. Avila, E. Frantzeskakis, A. Levy and M.C. Asensio, Nature Commun. 6 (2015) 8585.

\bibitem{landau} L.D. Landau, Phys.Z.Sowjetunion 3 (1933) 644.
 
\bibitem{chadan-khuri-martin-wu} K. Chadan, N.N. Khuri, A. Martin and T.T. Wu, Phys. Rev. D58 (1998) 025014; J. Math. Phys. 44 (2003) 406.

\bibitem{kato} T. Kato, ``Perturbation theory for linear operators'' (Springer-Verlag, Heidelberg, 1976).

\bibitem{newton-seto} N. Set\^o, Publ. of RIMS 9 (1974) 429; B. Simon, Ann. Phys. (NY) 97 (1976) 279; R.G. Newton, ``Scattering Theory of Waves and Particles'', (Springer-Verlag, New York, 1982).

\bibitem{bargmann} V. Bargmann, Proc. Nat. Acad. Sci. U.S.A. 38 (1952) 961; B. Simon, ``On the number of bound states of two-body Schr\"odinger operators: A review'', in Studies in Mathematical Physics, ``Essays in Honour of Valentine Bargmann'' (Princeton University Press, Princeton, 1976) 305.
 
\bibitem{binegar} B. Binegar, J. Math. Phys. 23 (1982) 1511.

\bibitem{sakurai} J.J. Sakurai, ``Advanced Quantum Mechanics'', Addison-Wesley, 1967; C. Itzykson and J.-B. Zuber, ``Quantum Field Theory'', McGraw-Hill, 1988.

\bibitem{del_cima-miranda} E.S. Miranda, ``Electron-electron attractive potential in parity-preserving Maxwell-Chern-Simons QED$_3$, M.Sc. Thesis - UFV - (2016); O.M. Del Cima and E.S. Miranda, ``On the emergence of electron-electron attractive interaction in parity-preserving $U(1)\times U(1)$ massive QED$_3$'', in progress.

\bibitem{froissart-martin-bound} M. Chaichian, J. Fischer, Yu.S. Vernov, Nucl. Phys. B383 (1992) 152; O.M. Del Cima, Mod. Phys. Lett. A9 (1994) 1695; K. Chadan, N.N. Khuri, A. Martin, T.T. Wu, Phys. Rev. D58 (1998) 025014.

\bibitem{del_cima-franco-lima} W.B. De Lima, O.M. Del Cima and D.H.T. Franco, ``An estimate for the number of bound states in planar quantum field theories'', in progress.

\bibitem{pair-coherence-length} M. Kim, D. Jeong, G.-H. Lee, Y.-S. Shin, H.-W. Lee and H.-J. Lee, Sci. Rep. 5 (2015) 8715.

\bibitem{graphene-superconductor} S. Ichinokura, K. Sugawara, A. Takayama, T. Takahashi and S. Hasegawa, ACS Nano 10 (2016) 2761; Y. Cao, V. Fatemi, S. Fang, K. Watanabe, T. Taniguchi, E. Kaxiras and P. Jarillo-Herrero, Nature 556 (2018) 43; Y. Cao, V. Fatemi, A. Demir, S. Fang, S.L. Tomarken, J.Y. Luo, J.D. Sanchez-Yamagishi, K. Watanabe, T. Taniguchi, E. Kaxiras, R.C. Ashoori and P. Jarillo-Herrero, Nature 556 (2018) 80.
\end{references}
\end{document}